\documentclass[letter]{aa}  

\usepackage{graphicx}
\usepackage{txfonts}
\usepackage{hyperref}
\usepackage{nicefrac, xfrac}
\newcommand{\Msun}{M$_{\odot}$}

\usepackage{supertabular,booktabs}
\usepackage{tabularx}
\usepackage{longtable}

\begin{document}

   \title{A neutron star candidate in the long-period binary 56 UMa}

   \author{A. Escorza\inst{1}
          \and
          D. Karinkuzhi\inst{2}
          \and
          A. Jorissen\inst{3}
          \and
          S. Van Eck\inst{3}
          \and
          J. T. Schmelz\inst{4}
          \and
          G. L. Verschuur
          \and
          H. M. J. Boffin\inst{5}
          \and
          R. J. De Rosa\inst{1}
          \and
          H. Van Winckel\inst{6}
          }

   \institute{European Southern Observatory, Alonso de C\'{o}rdova 3107, Vitacura, Santiago, Chile. \email{ana.escorza@eso.org}
   \and
   Department of Physics, University of Calicut, Thenhipalam, Malappuram, 673635, India
   \and
   Institut d’Astronomie et d’Astrophysique, Université Libre de Bruxelles, ULB, Campus Plaine C.P. 226, Boulevard du Triomphe, B-1050, Bruxelles, Belgium
   \and
   USRA, 7178 Columbia Gateway Drive, Columbia, 21046, MD
   \and
   European Southern Observatory, Karl-Schwarzschild-Strasse 2, 85748, Garching bei M\"{u}nchen, Germany
   \and
   Institute of Astronomy, KU Leuven, Celestijnenlaan 200D, B-3001, Leuven, Belgium
   }

   \date{Received \today}

% \abstract{}{}{}{}{} 
% 5 {} token are mandatory

  \abstract
   {56 UMa is a wide binary system that contains a chemically peculiar red giant and a faint companion. Due to its surface chemical abundances, the red giant was classified as a barium (Ba) star. This implies that the companion has to be a white dwarf, since Ba stars form when mass is transferred to them from an s-process rich Asymptotic Giant Branch (AGB) star. However, in the case of 56~UMa, the companion might be too massive to be the progeny of an AGB star that efficiently produced s-process elements such as barium.}
   {In this Letter, we revisit the orbital parameters of the system and perform a full spectral analysis with the goal of investigating the Ba-star classification of the giant and unravelling the nature of its faint companion.}
   {We combined radial-velocity and astrometric data to refine the orbital parameters of the system, including the orbital inclination and the companion mass. Then, we re-determined the stellar parameters of the giant and its chemical abundances using high-resolution HERMES spectra. Finally, we investigated the morphology of the interstellar gas in the vicinity of the system.}
   {The faint component in 56~UMa has a mass of $1.31\pm0.12$~\Msun, which, together with the mixed s+r abundance profile of the red giant, confirms that the giant is not a standard barium star. Additionally, the clear identification of a cavity surrounding 56~UMa could indicate that a supernova explosion occurred about 10$^{5}$ years ago in the system, suggesting that the faint companion might be a neutron star. However, finding an evolutionary scenario that explains all the observables is not trivial, so we discuss different possible configurations of the system and their respective merits.}
   {}


   \maketitle
%

%--------------------------------------------------------------------
\section{Introduction}\label{intro}

56 Ursae Majoris (also known as 56~UMa, HD\,98839, or HIP\,55560) is a binary system that contains a G8-type giant and a faint lower-mass companion \citep{Keenan89}. The giant was identified as a mild barium (Ba) star by \cite{Keenan77}. Ba stars are chemically peculiar stars that show an overabundance of carbon and elements formed by the slow-neutron-capture (s-) process of nucleosynthesis \citep{BidelmanKeenan51}, which mostly operates in the interiors of Asymptotic Giant Branch (AGB) stars \citep[e.g.][]{Lugaro03,Cristallo09,Karakas10}. Ba stars form when an AGB star transfers enriched material to a companion in a binary system \citep{McClure80}. According to this scenario, the donors in these systems evolved off the AGB long ago and are now dim white dwarfs (WDs).

Even though the Ba-star classification of the giant star in 56~UMa has been debated \citep[e.g.][]{Keenan89, Lu91, Griffin1996, Karinkuzhi18}, and its high luminosity and mass made it an outlier among Ba giants \citep{Jorissen19}, the nature of its companion has not been extensively discussed. \cite{Bohm-Vitense84} noticed that the International Ultraviolet Explorer (IUE) spectrum of 56~UMa showed two weak emission lines at 147.4 and 166.8~nm, likely due to the N IV] and O III] semi-forbidden lines. They proposed that these lines arise from an extended region of high-temperature, low-density gas coming from a small, hot source in the system, and suggested that this source might be a massive WD. However, it is interesting to note that the same lines arise in the hot gas surrounding accreting neutron stars \citep[NSs;][]{Anderson1994,CastroSegura2022}.

\begin{figure*}[t]
\centering
\includegraphics[width=0.9\textwidth]{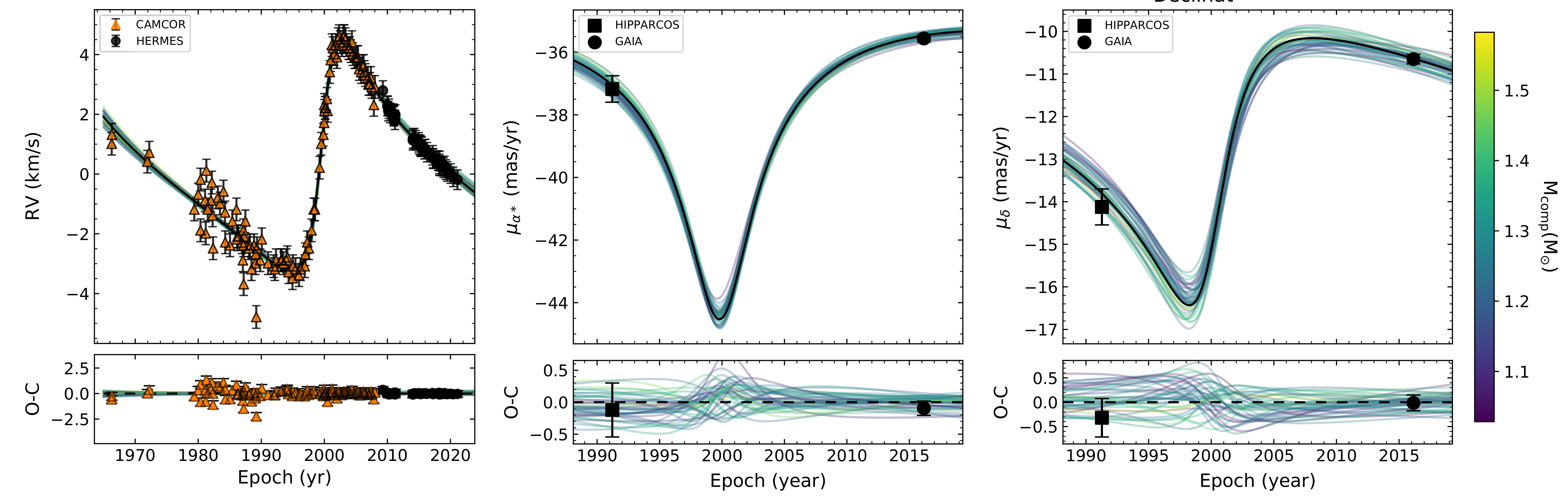}
\caption{Fits to the RV curve (left) and the \textit{Hipparcos} and \textit{Gaia} proper motions (middle and right) of 56~UMa. The black line is the orbit with the maximum likelihood, and 40 additional well-fitting solutions are included, colour-coded as a function of companion mass.}\label{fits}
\end{figure*}

At the location of 56~UMa, there is also a strong detection in the GALEX (Galaxy Evolution Explorer) far-UV band (153~nm), at the level of $7.66 \times 10^{-5}$~Jy (GALEX-DR5; \citealt{Bianchi2011}). This is comparable with the $6.09 \times 10^{-6}$~Jy measured at 212~nm (XMM-SUSS5.0; \citealt{Page2012}) for the closest (120~pc) NS to Earth, RX J1856.5-3754 \citep{Walter2010}. 56~UMa is located at 170~pc \citep{GaiaEDR3}. Finally, in a study of coronal X-ray emission in late-type giants, \cite{Gondoin1999} reports an X-ray luminosity of $L_X = 1.07 \times 10^{29}$~erg~s$^{-1}$ for 56~UMa. This could be ascribed to coronal emission or to X-ray emission associated with a NS. Neutron stars are observed to have masses between 1.2 and 2.35~\Msun\ \citep{Romani22} and typically emit at radio and X-ray wavelengths when they are young and rapidly rotating. As they age, their rotation slows down and their emission fades, unless they are fuelled by mass transfer from a companion. Neutron stars not powered by accretion are difficult to detect and, therefore, very rarely observed \citep{Mazeh2022}.

Here, we revisit the stellar and orbital parameters of 56~UMa (Sect. \ref{orbit}), the stellar properties of the giant component (Sect. \ref{m1}), and its heavy-element abundances (Sect. \ref{abundances}). We show possible evidence of a supernova explosion in the system (Sect. \ref{sec:cavity}), and we discuss different evolutionary scenarios compatible with these observations (Sect. \ref{sec:discussion}), including the possibility that the faint component of 56~UMa is a NS.

%--------------------------------------------------------------------
\section{Orbital parameters and companion mass}\label{orbit}

In order to obtain the orbital parameters of 56~UMa, we fit a Keplerian orbital model to the available radial-velocity (RV) and astrometric data. We used CAMCOR (Coravel
at the Cambridge 36-inch telescope; \citep{Griffin08}) and HERMES (High-Efficiency and high-Resolution Mercator Echelle Spectrograph; \citealt{Raskin11,Jorissen19}) RV data, the positions and proper motions provided by the \textit{Hipparcos} catalogue \citep{ESA1997, HippCat97} and the \textit{Gaia} Early Data Release 3 \citep[EDR3;][]{GaiaEDR3summary, GaiaEDR3}, the individual astrometric measurements from the re-reduction of the \textit{Hipparcos} intermediate astrometric data \citep{vanLeeuwen07} and the \textit{Hipparcos}-\textit{Gaia} Catalogue of Accelerations (HGCA; \citealt{Brandt18, HGCA-EDR3}). We used the code \textsc{orvara}, developed by \cite{Brandt21}, to exploit the combinations of these datasets. \textsc{orvara} fits a single Keplerian model to the datasets described above, employing a parallel-tempering Markov chain Monte Carlo minimisation routine \citep[\textsc{ptmcmc};][]{PTMCMC}. We fit ten parameters: the six Keplerian orbital elements, the masses of the two components and a RV jitter per instrument to be added to the RV uncertainties. We assumed uninformative priors for the orbital elements and the faint companion mass, but we adopted a Gaussian prior of $4.3\pm 0.6$~\Msun\ \citep{Jorissen19} for the primary mass (we used a 3$\sigma$ uncertainty to account for systematic errors in the mass determination) and another Gaussian prior for the parallax that corresponded to the EDR3 value. The results are presented in Table \ref{table:MCMCresults}.

\renewcommand{\arraystretch}{1.2}
\begin{table}[t]
\begin{center}
\caption{Stellar and orbital parameters of the 56~UMa system}\label{table:MCMCresults}
\vspace{1mm}
\begin{small}
\begin{tabular}{l c}
\hline
{\bf Parameter} & {\bf Median $\pm$ 1$\sigma$}\\
\hline
Temperature, $T_{\rm eff}$ [K] & 4917 $\pm$ 34\\
Surface gravity, log \textit{g} & 2.3 $\pm$ 0.6\\
Metallicity, [Fe/H] & $-0.05$\\
Microturbulence, $\xi$ [km/s] & 1.56\\
Primary mass, $M_{\rm 1}$ [$M_{\odot}$] & 4.3 $\pm$ 0.2 \\
\hline
Period, $P$ [days] & 16911$^{+438}_{-401}$\\
Eccentricity, $e$ & 0.562$^{+0.012}_{-0.012}$\\
Semi-major axis, $a$ [AU] & 22.9$^{+1.0}_{-1.1}$\\
Argument of periastron, $\omega_{\rm 1}$ [$^{\circ}$] & 286$^{+2.3}_{-2.3}$\\
Time of periastron, $T_{0}$ [HJD] & 2,468,401$^{+432}_{-385}$\\
Parallax, $\varpi$ [mas] & 5.86$^{+0.03}_{-0.04}$ \\
Ascending node, $\Omega$ [$^{\circ}$] & 60$^{+3}_{-3}$\\
Inclination, $i$ [$^{\circ}$] & 68$^{+3.6}_{-3.4}$ \\
Secondary mass, $M_{\rm 2}$ [$M_{\odot}$] & 1.31$^{+0.11}_{-0.12}$\\
Center-of-mass velocity [km/s] & $0.13\pm 0.01$\\
Center-of-mass $\mu_{\alpha*}$ [mas/yr] & $-37.32\pm 0.01$\\
Center-of-mass $\mu_{\delta}$ [mas/yr] & $-12.18\pm 0.01$\\
\hline
\end{tabular}
\end{small}
\end{center}
\end{table}

\begin{figure*}[t]
\centering
\includegraphics[width=0.7\textwidth]{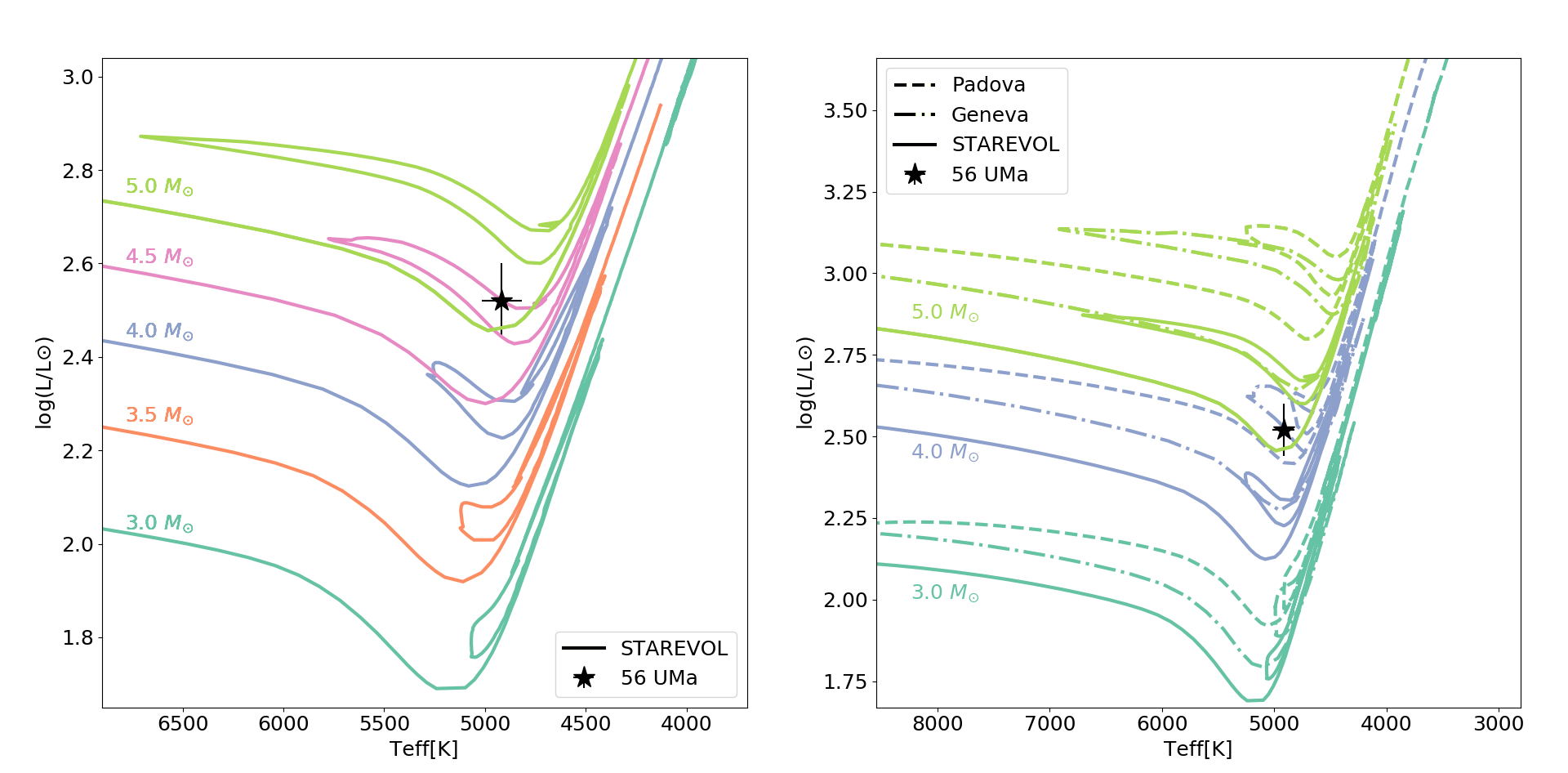}
\caption{Location of 56~UMa in the HRD. On the left, we included STAREVOL solar-metallicity tracks, and on the left, we compared STAREVOL (solid lines), Geneva (dot-dashed lines) and Padova (dashed lines) tracks for 3, 4 and 5~M$_\odot$ stars.}\label{hrd}
\end{figure*}

Figure \ref{fits} shows the fit to the RV data and to the \textit{Hipparcos} and \textit{Gaia} proper motions, and Fig. \ref{fig:corner} shows a corner plot with some derived parameters. The system is in an orbit of about 46 years and has a significant eccentricity of 0.56. Our joint RV and astrometric fit yields orbital parameters that are compatible with the results published by \cite{Jorissen19}. Additionally, for the first time, we constrained the inclination to $67.5 \pm 4^{\circ}$ and the secondary mass to $1.31 \pm 0.12$~M$_{\odot}$. This mass estimate, which is compatible with the value estimated by \cite{Kervella19} from the proper motion anomaly only, raises three issues: (i) the companion mass is surprisingly high for WDs and close to the Chandrasekhar mass \citep{Chandrasekhar1939}; (ii) in the context of Ba stars, this WD would be significantly more massive than any other Ba-star companion (see \citealt{Escorza23}); and (iii) the AGB progenitor of such a WD would be at the upper end of intermediate-mass stars, where the s-process is predicted to be fairly inefficient \citep{Kappeler11}. Alternatively, this high mass is compatible with the mass of a NS.

%--------------------------------------------------------------------
\section{Constraining the primary mass}\label{m1}

Given the strong correlation between the primary and secondary masses (see Fig. \ref{fig:corner}) and the fact that we used a prior for the primary mass, we re-evaluated this value. To do this, we co-added 27 high-resolution (R $= 86\,000$) HERMES \citep{Raskin11} spectra to first recalculate the stellar parameters of the giant.
The signal-to-noise ratio (S/N) of the co-added spectrum was about 500, and the stellar parameters were derived using the Brussels Automatic Code for Characterizing High accUracy Spectra (BACCHUS) in a semi-automated mode \citep{Masseron2016}. BACCHUS uses interpolated MARCS model atmospheres \citep{Gustafsson2008} and the 1D local thermodynamical equilibrium (LTE) spectrum-synthesis code TURBOSPECTRUM \citep{Plez2012}. It computes abundances using equivalent widths or spectral synthesis, allowing excitation and ionization equilibria to be checked for, thereby constraining $T_{\rm eff}$ and $\log$ g. It also calculates the microturbulence velocity ($\xi$) by ensuring consistency between the Fe abundances derived from lines of various equivalent widths. We adopted solar abundances from \cite{Asplund2009} and the line lists assembled in the framework of the \textit{Gaia}-ESO survey \citep{Heiter2015, Heiter2020}.

Our analysis confirms the stellar parameters derived by \cite{Jorissen19} and the updated results are listed in Table \ref{table:MCMCresults}. Our results are also in agreement with those previously derived by other authors (Table~\ref{tab:atmospheric_parameters}). A luminosity of $L = 332$~L$_\odot$ was also derived by \cite{Jorissen19} by integrating the spectral energy distribution of the star and using the \textit{Gaia} Data Release (DR) 2 parallax ($5.87\pm0.19$~mas; \citealt{GaiaDR2}), following \cite{Escorza17}. The \textit{Gaia} DR2 parallax is in agreement with its re-evaluation in DR3 ($5.88\pm0.093$~mas, \citealt{GaiaEDR3}). With the latter as an input, our Markov chain Monte Carlo (MCMC) optimisation converges to $5.86\pm0.04$~mas (Table \ref{table:MCMCresults}), so the luminosity quoted above remains valid. 

Once $T_{\rm eff}$ and $L$ are confirmed, the evolutionary mass of the giant can be revisited. \cite{Jorissen19} obtained it by comparing its location in the Hertzsprung-Russell diagram (HRD) with solar-metallicity tracks from STAREVOL \citep{Siess00, Siess06}. However, evolutionary tracks differ from one another depending on the prescriptions used for overshooting, rotation, mass loss, and so on. The right panel of Fig.~\ref{hrd} shows a comparison between STAREVOL, Geneva \citep{Ekstrom12} and Padova \citep{Girardi00} models of solar metallicity ($Z_{\odot} = 0.0134, 0.014$, and $0.019$, respectively) and demonstrates that a range from 4 to 5~\Msun\ is suitable for the red giant, independently of the used models. That range for the primary mass implies a range from 1.1 to 1.5 \Msun\ for the companion.

%--------------------------------------------------------------------
\begin{figure*}[t]
\centering
\includegraphics[width=0.8\textwidth]{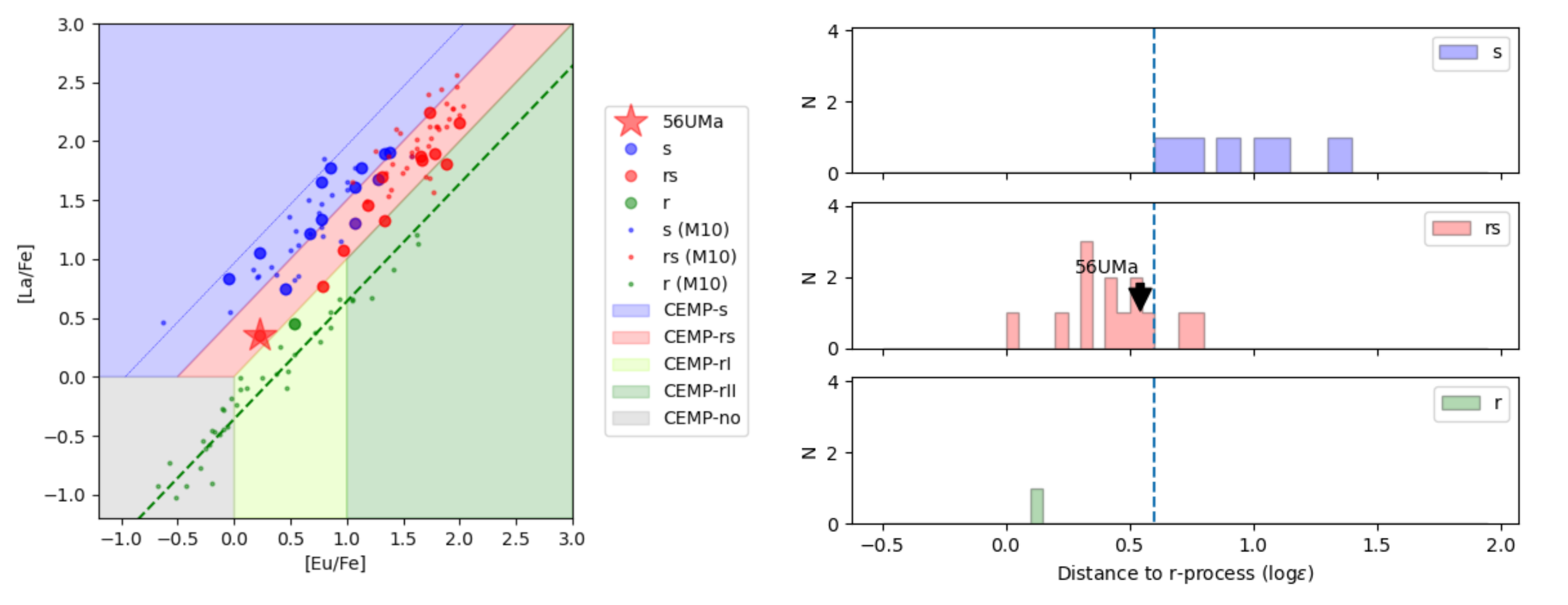}
\caption{56~UMa compared to the CEMP stars from \cite{Karinkuzhi-2021}. Left: [La/Fe]-[Eu/Fe] plane, where the dashed green line corresponds to a pure solar r-process, and the continuous blue line to s-process predictions. Right: Histograms of the abundance distance $d$ (Appendix \ref{sec:d}) of 56~UMa and the reference stars. The vertical dashed line at $d$ = 0.6 marks the threshold adopted between CEMP-s and CEMP-rs stars.
}\label{Fig:abundances}
\end{figure*}

\section{Chemical abundance profile of the red giant}\label{abundances}

To investigate the origin of the chemical enrichment of the giant, we derived its chemical abundances, including the abundances for 14 heavy elements, some of which are predominantly produced by the s-process, some by the r-process, and some by both. The atomic lines we used, including hyperfine (HF) splitting when available, are listed in Table~\ref{Tab:linelist}.

The abundances of four s-process elements (Y, Zr, La, and Ce) were published by \cite{Jorissen19}, but we present a new analysis that uses the co-added spectrum with a higher S/N. The measured abundances and the non-LTE corrections, when available, are listed in Table \ref{Tab:abundances}. A non-LTE correction of 0.19 dex, corresponding to the atmospheric parameters of the \ion{Sr}{i} line, was adopted from \cite{Bergemann2012}. Table \ref{Tab:uncertainties} lists the sensitivity of the abundances to variations in the atmospheric parameters (determined following \citealt{Karinkuzhi18}). Figure \ref{Fig:HD98839_synth} shows the fit to a few heavy-element lines.

56~UMa has a (moderate) enhancement in s-process elements, consistent with its mild Ba-star classification, but also an unexpected r-process enhancement. A small group of carbon-enriched metal-poor (CEMP) stars are known to have such a hybrid (s+r) profile \citep{Karinkuzhi-2021}, but contrarily to these CEMP-rs stars, 56~UMa has a solar metallicity and a solar C/O ratio. The left panel of Fig. \ref{Fig:abundances} shows the location of 56~UMa in the [La/Fe]-[Eu/Fe] plane compared to the CEMP stars studied by \cite{Karinkuzhi-2021}. Additionally, we computed an abundance distance, $d$, from the solar r-process abundance pattern, which is described in Appendix \ref{sec:d}. The r-process was adopted as reference because, although it shows some variations \citep{Goriely97b,Roederer11}, it appears less sensitive to its operating conditions (metallicity, radiative or convective environment, mixing of protons into the C-rich intershell, etc.) than the s-process. The abundance distance, $d$, allowed us to discern whether the star abundance profile is pure r ($d < 0.25$), predominantly s ($d > 0.6$), or hybrid s+r ($0.25 < d < 0.6$) as defined by \cite{Karinkuzhi-2021}. 56~UMa, with $d=0.54$, falls among the CEMP-rs objects (right panel of Fig.~\ref{Fig:abundances}). \cite{Karinkuzhi-2021} showed that AGB nucleosynthesis may account for such a mixed s+r profile, but only at low metallicity. To our knowledge, 56~UMa is the first object at solar metallicity that shows an enrichment of both s- and r-process origin.

%--------------------------------------------------------------------
\section{A neutral hydrogen cavity}\label{sec:cavity}
\begin{figure}[t]
\centering
\includegraphics[width=0.4\textwidth]{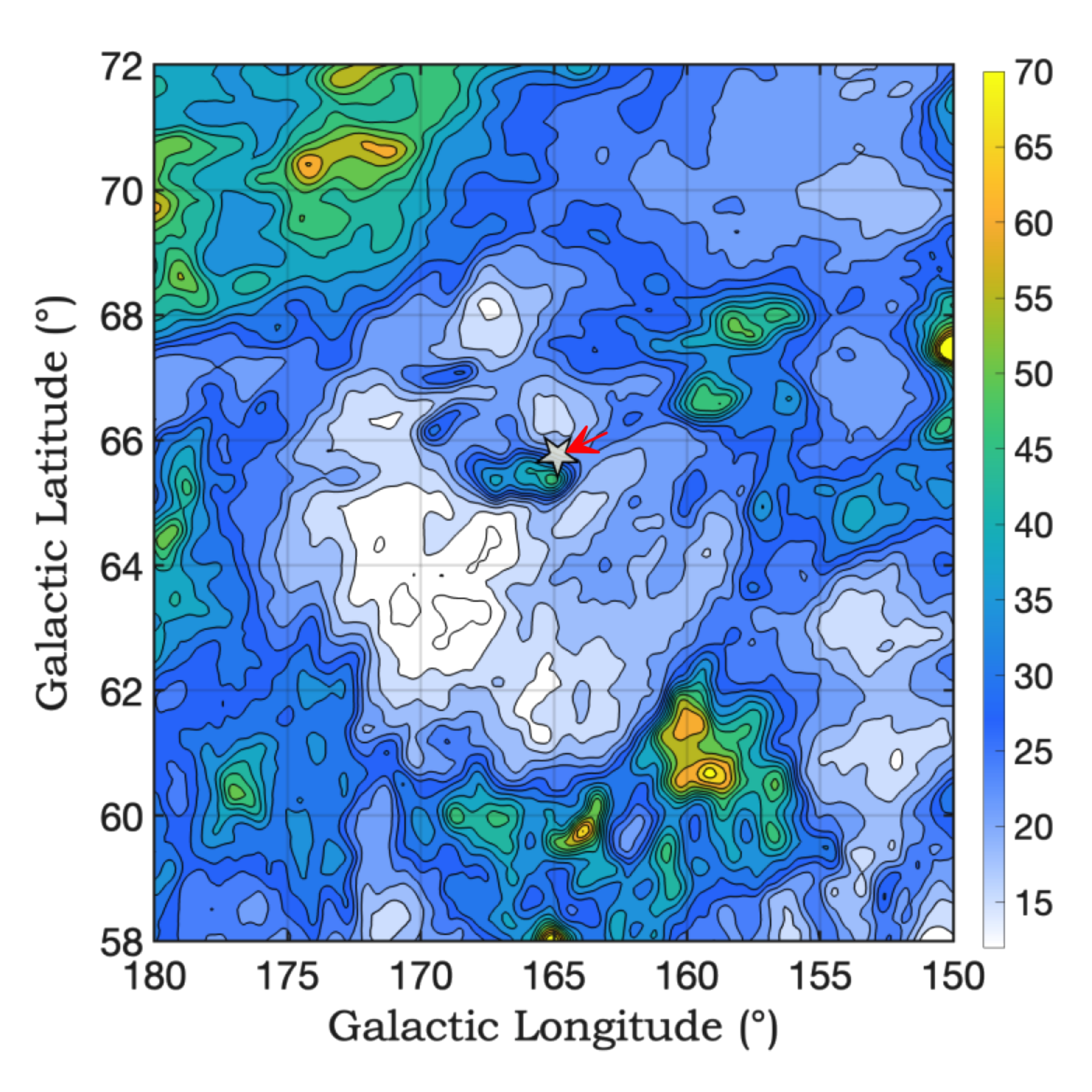}
\caption{Galactic-coordinate map of H\textsc{i} at $\lambda$ 21 cm integrated between $\pm$ 25 km s$^{-1}$ and centred on 56 UMa. The arrow corresponds to the motion of the star over the last $10^5$~yr \citep{Schmelz2022}, and was derived from the orbital solution (Table~\ref{table:MCMCresults}). Colour bar units are kelvins.}\label{cavity}
\end{figure}

We also investigated the morphology of the interstellar gas in the vicinity of 56~UMa. Figure \ref{cavity} shows an integrated $\lambda$ 21 cm map of neutral atomic hydrogen in Galactic coordinates, centred on 56~UMa for H\textsc{i} with velocities between $\pm$~25~km~s$^{-1}$ (the center-of-mass velocity of 56~UMa is 0.13 km/s; Table~\ref{table:MCMCresults}). The data are from the HI4PI survey \citep{Bekhti16}, which has an angular resolution of $\Theta_{\rm FWHM}=16^{\prime}2$, a sensitivity of $\sigma = 43$~mK, and full spatial sampling of 5$^{\prime}$ in both Galactic longitude and latitude.

While there is complex structure in the area, the image shows a clear absence of emission, a cavity, surrounding 56~UMa, although the star is not exactly at its centre. \cite{Schmelz2022} propose that this cavity is associated with the remnant of a SN that exploded about $10^5$~years ago. Its rim is steepest in the lower-right quadrant, where we have the clearest view of the material that may have piled up during the snow-plough phase. The minimum column density is found at Galactic coordinates $l, b = 170.67, 62.42$ is 6.5 $\times$ 10$^{18}$ cm$^{-2}$, and there is also about 1.2\Msun\ \citep{Schmelz2022} of low-velocity gas in the cavity that could be associated with mass lost by the giant after the explosion. The clear identification of this cavity is only possible because it is within the Local Chimney \citep{Lallement03}, a low-density extension of the Local Bubble \citep{Zucker22} that stretches from the solar neighbourhood to the Galactic halo.

The unseen companion of 56~UMa has traditionally been identified as a WD, but if it were a NS, it could be the remains of the SN that evacuated the MI cavity and blasted MI itself outwards at 120~km/s \citep{Schmelz2022}. Then, one might expect to find evidence for high energy emission, as was indeed found by ROSAT as diffuse \nicefrac{1}{4}~keV X-rays in the lower-left half of the cavity, the area of the lowest H\textsc{i} emission \citep{Herbstmeier95,Schmelz2022}.

%--------------------------------------------------------------------
\section{Summary, discussion and conclusions}\label{sec:discussion}
We have presented the orbital, stellar, and chemical properties of the wide binary system 56~UMa and determined the mass of the unseen component to be 1.31$\pm$0.12~\Msun, which is compatible with both a WD and a NS. In the remainder of this section, we discuss the possible configurations of the system along with their pros and cons.

\begin{enumerate}
\item The faint companion could be a 1.3~\Msun\ WD, implying an AGB progenitor of a rather high mass, of the order of 6~\Msun\ \citep{Marigo22}. However, massive AGB stars are not expected to be efficient enough s-process producers \citep{Kappeler11} to account for the observed barium overabundance. Additionally, the lack of rubidium overabundance, which is considered the major signature of super-AGB nucleosynthesis \citep{GarciaHernandez06, Doherty17}, is also an argument against this scenario. A new exploration of the nucleosynthesis parameter space could add to this discussion. Finally, this scenario would require the observed H\textsc{i} cavity to have been created by an AGB (super-)wind or by a last-thermal pulse ejection event that produced a shock wave that cleared out the accumulated material around the AGB star. However, we would then need to understand why not all AGB stars are surrounded by interstellar medium cavities. As already mentioned, this cavity was discovered only because it is in a low-density environment, and such favourable conditions probably do not occur for each AGB star. An alternative option is that the cavity is not associated with any of the components of the 56~UMa system, given its off-centre location.

\item The companion could be a close pair formed by two low-mass stars \citep[e.g.][]{vandenheuvel2020}. If they were both main-sequence stars, mass transfer could not explain the peculiar abundances of the giant. Therefore, at least one companion must be an evolved star, a WD. Assuming we can extrapolate the stability criterion discussed by \cite{Tokovinin14} to this more evolved system, the ratio between the two periods must be higher than 4.7. If 56~UMa were a triple system, the inner orbit would have to be shorter than 3600~days. Most Ba stars are in orbits below that threshold \citep{Jorissen19, Escorza19}, so that period should give enough space for an AGB star to evolve. A problem in this scenario is that this WD must still originate from a star more massive than the initial mass of the current primary, so we cannot circumvent the difficulties listed for the previous scenario.

\item The faint companion could be a NS. This scenario can explain the cavity better than the previous ones; however, the SN rate in the Milky Way is estimated to be one to two every 100~yr \citep[e.g.][]{Rozwadowska21}, so the probability of such an event occurring within 170~pc of Earth and within $10^{5}$~yr is very low. Even if we spotted such an event, the fact that the system is bound, yet very eccentric, together with the current location of the giant on the HRD require dynamical and evolutionary fine-tuning. We estimate that for this scenario to work, it would require an asymmetric explosion, with a small associated kick opposite to the orbital motion, in order to match the mass budget imposed by the evolutionary timescales (see Appendix \ref{sec:kick}). \cite{Schmelz2022} suggest that the SN exploded about $10^{5}$~yrs ago, and the Padova isochrones
\citep{Bressan2012} predict that a solar-metallicity 8~\Msun\ star would reach carbon ignition at an age of 40 Myr. At that age, a 4.3~\Msun\ star (the current mass of the giant) would still be on the main sequence, which is incompatible with its current status. A possible way to solve this tension is to assume that the initial masses of the two components were almost equal, and that the SN blast removed a couple of solar masses from the current red giant envelope. \cite{Leigh16} show, however, that red giant stars must be nearly completely stripped for their luminosity to change significantly. Hence, 56~UMa should have kept a luminosity typical of a 7-8 M$_{\odot}$ star, which it did not (Fig. \ref{hrd}). Additionally, numerical simulations of SN explosions in binaries \citep{Hirai2014} indicate that there is no mixing between the SN ejecta and the companion material, so the SN would leave no chemical imprint on the surviving companion. Thus, the chemical peculiarities detected on the giant remain unexplained in this scenario too.
\end{enumerate}

We conclude that none of the potential scenarios can explain all observables, and we cannot yet pin down the nature of the faint companion in 56~UMa. A crucial ingredient is whether or not the H\textsc{i} cavity is a SN remnant associated with 56~UMa. Even if it is not, the massive chemically peculiar giant in combination with the massive unseen companion is difficult to reconcile with stellar evolution.  We stress that the chemistry in 56~UMa, bearing imprints from both the s- and r-process of nucleosynthesis, is the first such star at solar metallicity. This system provides a strong test for binary evolution.\\

%--------------------------------------------------------------------
\begin{acknowledgements}
We thank N. Chamel, L. Siess and S. Goriely for the useful discussions. The HERMES spectrograph is supported by the Fund for Scientific Research of Flanders (FWO), the Research Council of KU Leuven, Belgium, the Fonds National de la Recherche Scientifique (F.R.S.-FNRS), the Royal Observatory of Belgium, the Observatoire de Genève, and the Thüringer Landessternwarte Tautenburg. This work makes use of SIMBAD, operated at CDS, Strasbourg, and of data from the European Space Agency (ESA), processed by the \textit{Gaia} Data Processing and Analysis Consortium (DPAC).
\end{acknowledgements}

\bibliographystyle{aa} % style aa.bst
\bibliography{references} % your references Yourfile.bib

\appendix

\section{MCMC corner plot}

\begin{figure*}[t]
\centering
\includegraphics[width=0.6\textwidth]{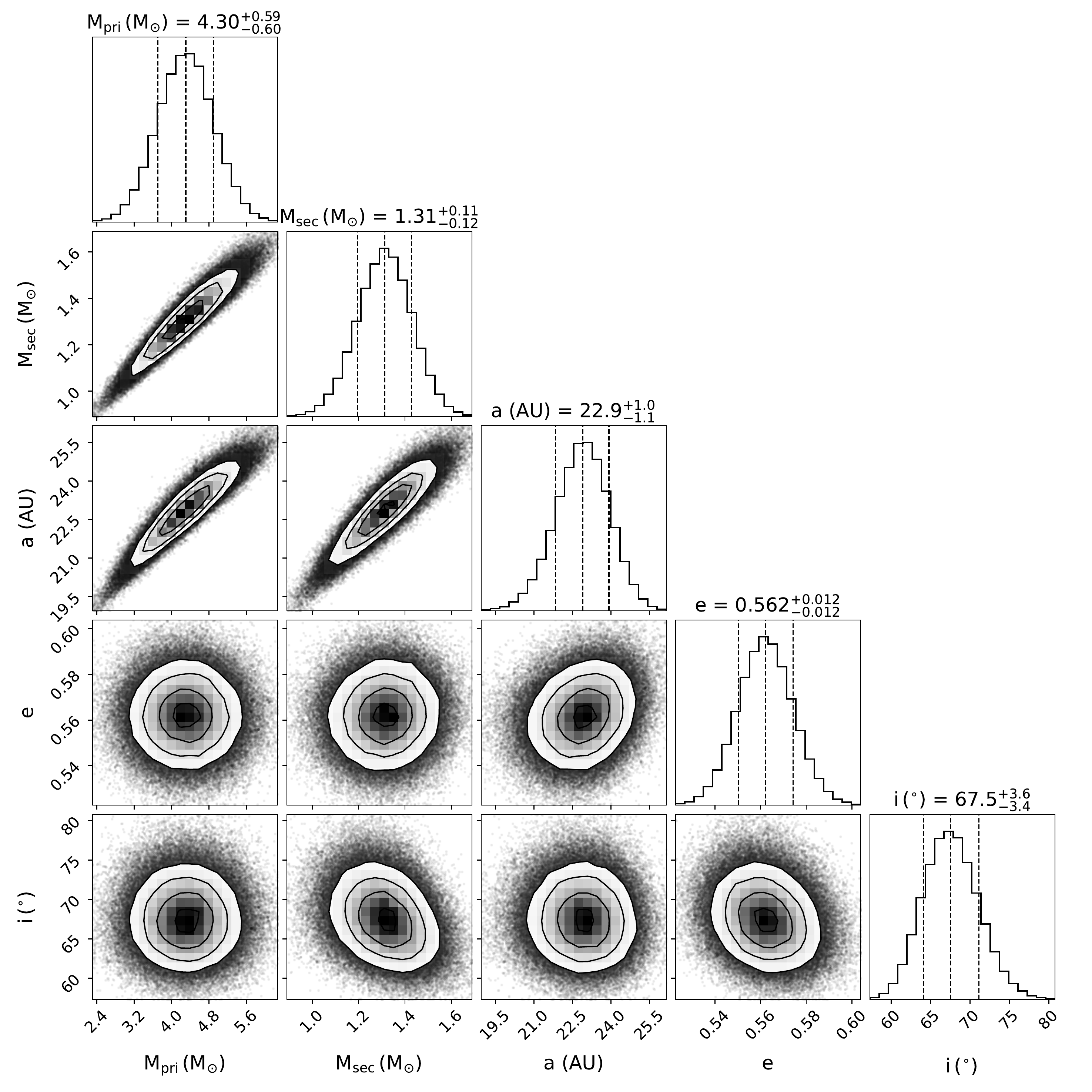}
\caption{MCMC corner plot.}\label{fig:corner}
\end{figure*}

Figure \ref{fig:corner} shows the 1D and 2D projections of the posterior probability distributions of the masses of the two stellar components in 56~UMa ($M_{\rm prim}$ corresponds to the red giant and $M_{\rm sec}$, to the faint companion) and a few orbital parameters (semi-major axis, $a$, eccentricity, $e$, and inclination, $i$) from the joint RV and astrometric MCMC fit (Sect. \ref{orbit}). The plot shows that the two masses are correlated, and that the semi-major axis is also correlated with the total mass of the system. 

\section{Literature parameters for 56~UMa}

We collected from the literature the atmospheric parameters derived for the giant star in 56~UMa over the past two decades. We list them in Table \ref{tab:atmospheric_parameters} ordered by decreasing $T_{\rm eff}$.

\begin{table}[t]
\centering
\caption{Atmospheric parameters (effective temperature, $T_{\rm eff}$, surface gravity, $\log g$, metallicity, [Fe/H] and microturbulence, $\xi$) for 56~UMa collected from the literature.} \label{tab:atmospheric_parameters}
\begin{tabular}{ccccl}
\hline
$T_{\rm eff}$ & $\log g$ & [Fe/H] & $\xi$ & Ref.\\
(K) &  &  &  (km/s)\\
\hline
 5088         & 2.84         & 0.24   & -& F16\\
 5043 $\pm$ 433 & 2.33         & 0.07   & -  & G09\\
 5010         & 2.1          & 0.0   & -  & L06\\
 4936 $\pm$ 25  & 2.3 $\pm$ 0.08 & $-0.05\pm0.04$ & 1.78 & T08\\
 4917 $\pm$ 34  & 2.3 $\pm$ 0.6  & $-0.05\pm0.12$ & $1.56\pm0.05$ & J19\\
% 4893         &              &                &  & 2005ApJ...626..446R \cite{Ramirez05}\\
%  4872         &              &                &SUN      & &1999A\&AS..139..335A \cite{Alonso99}\\
%  4890         & 2.55         &$-0.09$           &SUN     & &1990ApJS...74.1075M \cite{McWilliam90}\\
%  4900         & 2.00         & 0.23           &SUN      & &1990AJ.....99.1961F \cite{Fernandez-Villacanas90}\\
%  4893         & 2.43         & 0.06           &SUN      & &1981ApJ...247.1052S \cite{Sneden81}\\
% 4866         & 2.33         & 0.16           &SUN      & &1992A\&A...257..265L \AEnote{(not found)}\\
%  4846         & 2.0          &$-0.30$     &HD 113226    & &1977A\&A....54..465P \cite{Pilachowski77}\\
\hline\\
\end{tabular}
\textbf{Reference abbreviations:} F16: \cite{Feuillet16}; G09: \cite{GonzalezHernandez09}; L06: \cite{Lebre06}; T08: \cite{Takeda08}; J19: \cite{Jorissen19}.
\end{table}

\section{Spectral line list}

Table~\ref{Tab:linelist} presents the lines used in the present abundance analysis.\\
\vspace{5mm}
\topcaption{Spectral lines used in the abundance analysis.\label{Tab:linelist}}
\begin{center}
\begin{supertabular}{llr}
\hline
$\lambda$ & $\chi_{\rm low}$ & $\log gf$ \\
(\AA) & (eV) & \\
\hline\\
\tablehead{
\hline\\
$\lambda$ & $\chi_{\rm low}$ & $\log gf$ \\
(\AA) & (eV) & \\
\hline\\}

O I&      &      \\
6300.304& 0.000 &$-$9.715\\
7771.941&9.146&0.369\\
7774.161&9.146&0.223\\
7775.338&9.146&0.001\\
\hline\\
Na I&&\\
5682.633 & 2.102 &$-$0.706\\
5688.205 & 2.104 &$-$0.450\\
6154.226& 2.102 &$-$1.547 \\
6160.747&  2.104 &$-$1.246 \\
\hline\\
Mg I &&\\
4571.096 &0.000   &$-$5.623    \\
5528.405 &4.346   &$-$0.620    \\
\hline\\
Sr I     &            &         \\
4607.327 &0.000   & $-$0.570   \\
\hline\\
Y II     &        &         \\
4900.124 &  1.033 &   0.103 \\
5087.416 &  1.084 &  $-$0.170 \\
5200.406 &  0.992 &  $-$0.570 \\
5289.815 &  1.033 &  $-$1.850 \\
5320.782 &  1.084 &  $-$1.950 \\
5544.611 &  1.738 &  $-$1.090\\
5728.890 &  1.839 &  $-$1.120 \\
\hline\\
Zr I     &        & \\
6127.475 & 0.154 & $-$1.060  \\
6134.585 & 0.000 & $-$1.280  \\
6143.252 & 0.071 & $-$1.100  \\
\hline\\
Zr II     &        &        \\
4379.742  & 1.532  & $-$0.356 \\
5112.270  & 1.665  & $-$0.850 \\
\hline\\
Ba II  &&\\
4524.925 & 2.512 &$-$0.390\\
5853.669 & 0.604 &$-$1.967  \\
5853.669 & 0.604 &$-$1.967  \\
5853.670 & 0.604 &$-$1.909  \\
5853.670 & 0.604 &$-$2.113  \\
5853.671 & 0.604 &$-$1.909  \\
5853.671 & 0.604 &$-$2.113  \\
5853.672 & 0.604 &$-$2.113  \\
5853.672 & 0.604 &$-$2.511  \\
5853.673 & 0.604 &$-$0.909  \\
5853.673 & 0.604 &$-$0.909  \\
5853.673 & 0.604& $-$0.909  \\
5853.673 & 0.604&$-$1.812  \\
5853.673 & 0.604 &$-$2.113  \\
5853.673 & 0.604& $-$2.511  \\
5853.674  &0.604& $-$0.909  \\
5853.675 & 0.604& $-$0.909  \\
5853.675 & 0.604& $-$1.365  \\
5853.675 & 0.604& $-$1.812  \\
5853.675 & 0.604& $-$1.909  \\
5853.676 & 0.604& $-$1.365  \\
5853.676 & 0.604& $-$1.909  \\
5853.680 & 0.604 &$-$1.967  \\
5853.682 & 0.604 &$-$1.967 \\
\hline \\
La II     &        &        \\
4662.478  & 0.000  & $-$2.952 \\
4662.482  & 0.000  & $-$2.511 \\
4662.486  & 0.000  & $-$2.240 \\
4662.491  & 0.000  & $-$2.253\\
4662.492  & 0.000  & $-$2.137 \\
4662.493  & 0.000  & $-$2.256 \\
4662.503  & 0.000  & $-$2.511 \\
4662.505  & 0.000  & $-$2.056 \\
4662.507  & 0.000  & $-$1.763 \\
4748.726  & 0.927  & $-$0.540 \\
4920.965  & 0.126  & $-$2.261 \\
4920.965  & 0.126  & $-$2.407 \\
4920.966  & 0.126  & $-$2.065 \\
4920.966  & 0.126  & $-$2.078\\
4920.966  & 0.126  & $-$2.738 \\
4920.968  & 0.126  & $-$1.831 \\
4920.968  & 0.126  & $-$1.956 \\
4920.968  & 0.126  & $-$2.629 \\
4920.971  & 0.126  & $-$1.646 \\
4920.971  & 0.126  & $-$1.895 \\
4920.971  & 0.126  & $-$2.650 \\
4920.975  & 0.126  & $-$1.490 \\
4920.975  & 0.126  & $-$1.891 \\
4920.975  & 0.126  & $-$2.760\\
4920.979  & 0.126  & $-$1.354 \\
4920.979  & 0.126  & $-$1.957 \\
4920.979  & 0.126  & $-$2.972 \\
4920.985  & 0.126  & $-$1.233 \\
4920.985  & 0.126  & $-$2.162 \\
5290.818  & 0.000  & $-$1.650 \\
5301.845  & 0.403  & $-$2.587 \\
5301.857  & 0.403  & $-$2.684 \\
5301.860  & 0.403  & $-$2.508\\
5301.878  & 0.403  & $-$2.830 \\
5301.882  & 0.403  & $-$2.325 \\
5301.885  & 0.403  & $-$2.809 \\
5301.908  & 0.403  & $-$3.065 \\
5301.913  & 0.403  & $-$2.266 \\
5301.917  & 0.403  & $-$2.391 \\
5301.946  & 0.403  & $-$3.483 \\
5301.953  & 0.403  & $-$2.300 \\
5301.958  & 0.403  & $-$2.120 \\
5303.513  & 0.321  & $-$1.874 \\
5303.513  & 0.321  & $-$2.363 \\
5303.514  & 0.321  & $-$3.062 \\
5303.531  & 0.321  & $-$2.167 \\
5303.532  & 0.321  & $-$2.247 \\
5303.532  & 0.321  & $-$2.622 \\
5303.546  & 0.321  & $-$2.366\\
5303.546  & 0.321  & $-$2.622 \\
5303.547  & 0.321  & $-$2.351 \\
6262.113  & 0.403  & $-$3.047 \\
6262.114  & 0.403  & $-$2.901 \\
6262.132  & 0.403  & $-$2.705 \\
6262.134  & 0.403  & $-$2.718 \\
6262.135  & 0.403  & $-$3.378 \\
6262.164  & 0.403  & $-$2.471 \\
6262.166  & 0.403  & $-$2.596\\
6262.169  & 0.403  & $-$3.269 \\
6262.208  & 0.403  & $-$2.286 \\
6262.212  & 0.403  & $-$2.535 \\
6262.215  & 0.403  & $-$3.290 \\
6262.266  & 0.403  & $-$2.130 \\
6262.271  & 0.403  & $-$2.531 \\
6262.275  & 0.403  & $-$3.400 \\
6262.338  & 0.403  & $-$1.994 \\
6262.343  & 0.403  & $-$2.597 \\
6262.348  & 0.403  & $-$3.612\\
6262.422  & 0.403  & $-$1.873 \\
6262.429  & 0.403  & $-$2.802 \\
6262.434  & 0.403  & $-$4.015 \\
6390.455  & 0.321  & $-$2.012 \\
6390.468  & 0.321  & $-$2.183 \\
6390.468  & 0.321  & $-$2.752 \\
6390.479  & 0.321  & $-$2.570 \\
6390.479  & 0.321  & $-$3.752 \\
6390.480  & 0.321  & $-$2.390 \\
6390.489  & 0.321  & $-$2.536\\
6390.489  & 0.321  & $-$3.334 \\
6390.490  & 0.321  & $-$2.661 \\
6390.496  & 0.321  & $-$3.100 \\
6390.497  & 0.321  & $-$2.595 \\
6390.498  & 0.321  & $-$3.079 \\
6390.502  & 0.321  & $-$2.954 \\
6390.503  & 0.321  & $-$2.778 \\
6390.506  & 0.321  & $-$2.857 \\
\hline\\
Ce II     &        &        \\
4515.848  & 1.058  & $-$0.240\\
4562.359  & 0.478  &  0.230 \\
4628.169  & 0.516  &  0.200 \\
4943.441  & 1.206  & $-$0.360 \\
5274.229  & 1.044  &  0.130 \\
5330.556  & 0.869  & $-$0.400 \\
5472.279  & 1.247  & $-$0.100 \\
5975.818  & 1.327  & $-$0.460 \\
\hline\\
Nd II     &        &       \\
4451.560  & 0.380  & 0.070  \\
4947.020  & 0.559  & $-$1.130 \\
4961.387  & 0.631  & $-$0.710 \\
5089.832  & 0.205  & $-$1.160 \\
5092.788  & 0.380  & $-$0.610 \\
5132.328  & 0.559  & $-$0.710 \\
5212.360  & 0.205  & $-$0.960 \\
5276.869  & 0.859  & $-$0.440 \\
5293.160  & 0.823  &  0.100 \\
5311.450  & 0.986  & $-$0.420\\
5319.810  & 0.550  & $-$0.140 \\
5356.967  & 1.264  & $-$0.280 \\
5361.165  & 0.559  & $-$1.480 \\
\hline\\
Pr II     &        &        \\
5220.108  & 0.796  &  0.298 \\
5259.728  & 0.633  &  0.114 \\
5322.772  & 0.483  & $-$0.141 \\
\hline\\
Sm II     &        &       \\
4519.630  & 0.544  & $-$0.350 \\
4566.200  & 0.333  & $-$0.590 \\
4577.688  & 0.248  & $-$0.650 \\
4615.440  & 0.544  & $-$0.690 \\
\hline\\
Eu II     &        &        \\
6437.601  & 1.320  & $-$0.960\\
6437.603  & 1.320  & $-$0.960 \\
6437.606  & 1.320  & $-$2.191 \\
6437.609  & 1.320  & $-$2.191 \\
6437.617  & 1.320  & $-$2.191 \\
6437.619  & 1.320  & $-$2.191 \\
6437.620  & 1.320  & $-$1.070 \\
6437.623  & 1.320  & $-$1.998 \\
6437.627  & 1.320  & $-$1.070 \\
6437.627  & 1.320  & $-$1.998 \\
6437.630  & 1.320  & $-$1.181\\
6437.633  & 1.320  & $-$1.956 \\
6437.633  & 1.320  & $-$1.956 \\
6437.633  & 1.320  & $-$1.998 \\
6437.635  & 1.320  & $-$1.287 \\
6437.635  & 1.320  & $-$2.010 \\
6437.635  & 1.320  & $-$2.206 \\
6437.637  & 1.320  & $-$1.377 \\
6437.637  & 1.320  & $-$1.428 \\
6437.637  & 1.320  & $-$2.010 \\
6437.639  & 1.320  & $-$2.206\\
6437.640  & 1.320  & $-$1.998 \\
6437.647  & 1.320  & $-$1.181 \\
6437.652  & 1.320  & $-$1.956 \\
6437.657  & 1.320  & $-$1.956 \\
6437.662  & 1.320  & $-$1.287 \\
6437.667  & 1.320  & $-$2.010 \\
6437.669  & 1.320  & $-$2.010 \\
6437.674  & 1.320  & $-$1.377 \\
6437.677  & 1.320  & $-$2.206 \\
6437.679  & 1.320  & $-$2.206\\
6437.682  & 1.320  & $-$1.428 \\
6645.055  & 1.380  & $-$1.823 \\
6645.057  & 1.380  & $-$0.516 \\
6645.058  & 1.380  & $-$3.466 \\
6645.061  & 1.380  & $-$0.516 \\
6645.067  & 1.380  & $-$1.823 \\
6645.070  & 1.380  & $-$0.592 \\
6645.073  & 1.380  & $-$1.628 \\
6645.075  & 1.380  & $-$3.466 \\
6645.077  & 1.380  & $-$3.149\\
6645.080  & 1.380  & $-$0.672 \\
6645.085  & 1.380  & $-$1.583 \\
6645.086  & 1.380  & $-$0.592 \\
6645.087  & 1.380  & $-$0.754 \\
6645.091  & 1.380  & $-$3.076 \\
6645.093  & 1.380  & $-$0.838 \\
6645.093  & 1.380  & $-$1.634 \\
6645.094  & 1.380  & $-$1.628 \\
6645.097  & 1.380  & $-$0.921 \\
6645.099  & 1.380  & $-$1.829\\
6645.100  & 1.380  & $-$3.244 \\
6645.101  & 1.380  & $-$3.149 \\
6645.108  & 1.380  & $-$0.672 \\
6645.116  & 1.380  & $-$1.583 \\
6645.123  & 1.380  & $-$3.076 \\
6645.123  & 1.380  & $-$3.076 \\
6645.127  & 1.380  & $-$0.754 \\
6645.134  & 1.380  & $-$1.634 \\
6645.140  & 1.380  & $-$3.244 \\
6645.141  & 1.380  & $-$0.838 \\
6645.148  & 1.380  & $-$1.829\\
6645.153  & 1.380  & $-$0.921\\
\hline\\
Gd II   &    &      \\
4251.731  &0.382      & $-$0.220     \\
\hline\\
Dy II    &    &      \\
4103.306  &0.103 &$-$0.380       \\
\hline\\
Hf  II   &    &      \\
3918.090 & 0.452 & $-$1.140\\
4093.150 & 0.452 &$-$1.150     \\
\hline\\
Os I   &    &      \\
4260.849 & 0.000 &$-$1.434 \\
\hline
\end{supertabular}
\end{center}

\section{Individual abundances and their uncertainties}

\begin{figure}
\centering
\includegraphics[width=0.49\textwidth]{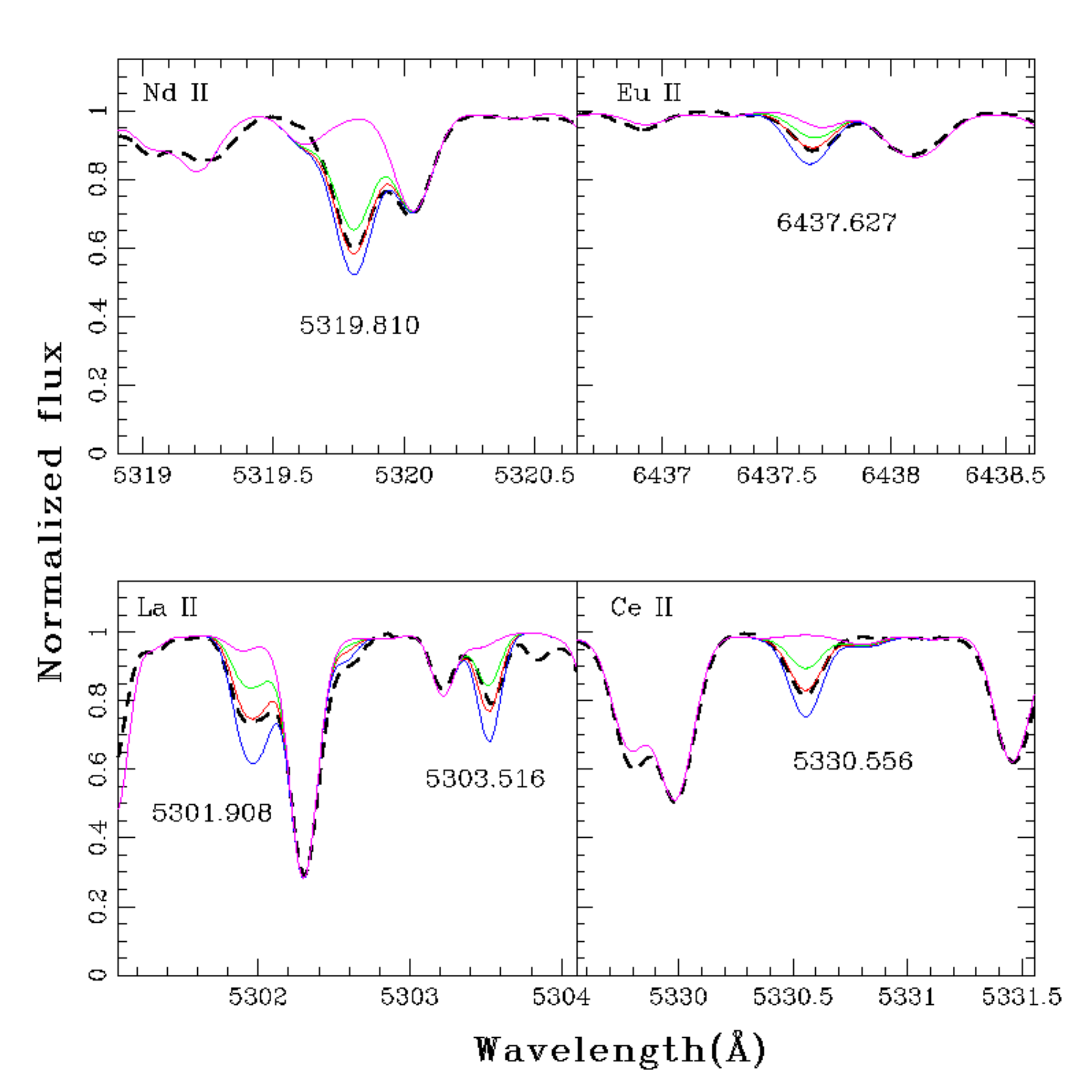}
\caption{Spectral fitting of \ion{La}{II}, \ion{Ce}{II}, \ion{Nd}{II} and \ion{Eu}{II}.  Red curves correspond to spectral syntheses with the adopted abundances for these elements, as listed in Table~\ref{Tab:abundances}. Blue and green curves correspond to syntheses with abundances deviating by  $\pm$ 0.3~dex from the adopted abundance.  The dashed black line represents the observed spectrum. The magenta line corresponds to the synthesis with a null abundance for the corresponding element.
\label{Fig:HD98839_synth} }
\end{figure}

\begin{table}[t]
\caption{Individual abundances of the giant component in 56~UMa}\label{Tab:abundances}
\begin{small}
\begin{center}
\begin{tabular}{lllccc}
\hline
\\
 &    $Z$  &    $\log \epsilon_{\odot}^{\rm a}$ & $\log {\epsilon}$ & $\sigma_{s}$ ($N$) & [X/Fe] $\pm\; \sigma_{\rm [X/Fe]}$\\
 %%%% HMJB %%%% Why is [XHe] not just the difference between the 2 log epsilon?
 %%%% AJ %%%% because [Fe/H] = -0.05
 &       &                         &                  &        &        \\
\hline
C$^{\rm b}$     &  6  & 8.43  &  8.20 & 0.06(4)    & $-$0.18$\pm$0.15 \\
$^{12}$C/$^{13}$C & &&  19 &         &                  \\
N$^{\rm c}$     &  7  & 7.83  &  8.40 & 0.09(30)&  0.62$\pm$0.21  \\
O$^{\rm d}$     &  8  & 8.69  &  8.70 & 0.00(2) & 0.06$\pm$0.22  \\
Na I  & 11  & 6.24  &  6.40& 0.10(4)    & 0.21$\pm$0.40 \\
Mg I  & 12  & 7.60  &  7.50:& 0.10(2)    &  $-$0.05$\pm$0.16  \\
Fe I  & 26  & 7.50  &  7.45    & 0.10(65)       &   --              \\
Rb I  & 37  & 2.52  & 2.50:   & 0.00(2) & 0.03$\pm$0.10   \\
Sr I  & 38  & 2.87  & 3.30:  & 0.10(1) & 0.48$\pm$0.35 \\
Sr I$_{\rm NLTE}$ & 38  & 2.87  & 3.49:  & 0.10(1) & 0.67$\pm$0.35 \\
%Sr I$_{0.283}$ & 38  & 2.87  & 2.40:  & 0.10(1) & $-$0.42$\pm$0.11 \\
%Sr I$_{NLTE_0.283}$ & 38  & 2.87  & 2.59:  & 0.10(1) & $-$0.23$\pm$0.11 \\
Y II  & 39  & 2.21  & 2.40  & 0.05(7) &  0.24$\pm$0.17 \\
Zr I  & 40  & 2.58  & 2.43  & 0.13(3) &  $-$0.10$\pm$0.30   \\
Zr II & 40  & 2.58  & 2.65  & 0.06(2) &  0.12$\pm$0.30  \\
%Nb I  & 41  & 1.46  &   & -   &  $\pm$0.16  \\
%Nb I  & 41  & 1.46  & -  & -   &  -         \\
Ba II & 56  & 2.18  & 2.72:  &   0.09(2)& 0.59$\pm$0.11  \\
%Ba II & 56  & 2.18  & 2.4:  &   0.10(1)& 0.27$\pm$0.10  \\
%Ba II (5853) & & 2.18  & 2.65  &   0.10(1)& 0.52$\pm$ 0.10 \\
La II & 57  & 1.10  & 1.40  & 0.12(8) &  0.35$\pm$0.22  \\
Ce II & 58  & 1.58  & 1.70  & 0.13(8) & 0.17$\pm$0.24  \\
Pr II & 59  & 0.72  & 0.75  & 0.05(3) & 0.08$\pm$0.25  \\
Nd II & 60  & 1.42  &1.72   & 0.13(13) & 0.35$\pm$0.22 \\
Sm II & 62  & 0.96  &1.08   & 0.13(4) &  0.17$\pm$0.32   \\
Eu II & 63  & 0.52  & 0.70  & 0.00(2) & 0.23$\pm$0.32  \\
Gd II & 64  & 1.10  & 1.40  &  0.10 (1)  & 0.35$\pm$0.21  \\
Dy II & 66  & 1.10  & 1.30  &  0.10 (1)  & 0.25$\pm$0.18  \\
Hf II & 72  & 0.85  & 1.10: & 0.00(2)    & 0.30$\pm$0.32\\
Os II & 76  & 1.40  & 1.60: &  0.10 (1)  & 0.25$\pm$0.20 \\
\hline
\end{tabular}\\
\end{center}
$^{\rm a}$ \cite{Asplund2009}\\
$^{\rm b}$ The carbon abundance was derived from the CH molecular bands at 4300~\AA\ and from the weak C$_2$ molecular lines at 5165~\AA\ and 5635~\AA.\\
$^{\rm c}$ The multiple CN bands longwards of 7000~\AA\ were used to derive the nitrogen abundance and the $^{12}$C/$^{13}$C ratio.\\
$^{\rm d}$ The oxygen abundance was derived from the [\ion{O}{I}] line at 6300~\AA and from the \ion{O}{I} resonance triplet at 7770~\AA. Due to non-LTE effects, the second abundance is higher by 0.3~dex. We took the average of these values after applying non-LTE corrections.
\end{small}
\end{table}

\begin{table}[t]
\begin{center}
\caption{Sensitivity of the abundances ($\Delta \log \epsilon_{X}$) to variations in the atmospheric parameters. }
\label{Tab:uncertainties}
\begin{small}
\begin{tabular}{crrrr}
\hline
\
       & \multicolumn{4}{c}{$\Delta \log \epsilon_{X}$} \\
        \cline{2-5}
 Element&   $\Delta T_{\rm eff}$ & $\Delta \log g$  &$\Delta$ [Fe/H]& $\Delta \xi_t$   \\
&   ($+$100 K) & ($+$0.5) & ($+$0.5  & ($+$0.5 \\
   &        &         & dex)& km~s$^{-1}$) \\
\hline\\
C  &$-$0.1   & 0.03 & 0.00   & 0.05  \\
N  & 0.20   &0.15   & 0.00   & 0.10	\\
O  & 0.00  & 0.20  & 0.00  & 0.30 	 \\
Na &0.35  & 0.25   & 0.25 & 0.25  \\
Mg &0.00  & 0.00   &0.00  & $-$0.20  \\
Fe &0.10  &0.06  & 0.00  & $-$0.03 \\
Rb &0.25   &0.00  &0.00  &0.00  \\
Sr & 0.40 &0.30  & 0.35  &  0.10  \\
Y  & $-$0.01 & 0.15  & $-$0.12  &0.06    \\
Zr &0.02  &0.30  & 0.05 & 0.20  \\
Ba &0.03  &0.15  &$-$0.20  &0.03\\
La &0.05   &0.20  &$-$0.01   & 0.15  \\
Ce &0.04  &0.28  & $-$0.12  & 0.20 \\
Pr &$-$0.03      &  0.25    &  0.02      & 0.17     \\
Nd & $-$0.03     & 0.20     &  $-$0.12      & 0.00     \\
Sm &  0.00    & 0.30     &  0.00     &  0.20    \\
Eu &  0.00    & 0.30     &  0.00      &  0.20    \\
Gd &   0.20   &  0.20    &  $-$0.05      &   0.20   \\
Dy &  0.00    & 0.00     &   $-$0.20     &  0.00    \\
Hf &   0.00   &   0.30   &    $-$0.10    & 0.15     \\
Os &0.30      &  0.10    &  $-$0.10      & 0.10     \\
\hline
\end{tabular}
\end{small}
\end{center}
\end{table}

Table \ref{Tab:abundances} lists the measured individual abundances along with non-LTE corrections when available, and Table \ref{Tab:uncertainties} lists the sensitivity of these abundances to variations in the atmospheric parameters. Figure \ref{Fig:HD98839_synth} shows the spectral fit for a few heavy-element lines.

The abundance uncertainties were calculated for all elements using the methodology described in \cite{Karinkuzhi-2021,Karinkuzhi2021b}. Following Eq.~2 from \cite{Johnson2002}, the uncertainties on the elemental abundances $\log \epsilon$ were calculated as:
\begin{equation}
\begin{split}
\label{Eq:Johnson}
\sigma^{2}_{\rm tot}=\sigma^{2}_{\rm ran}
\;+\; \left(\frac{\partial \log\epsilon}{\partial T}\right)^{2}\sigma^{2}_{T}  \;+\; \left(\frac{\partial \log \epsilon}{\partial \log g}\right)^{2}\;\sigma^{2}_{\log g} \\
\;+\; \left(\frac{\partial \log \epsilon}{\partial \xi }\right)^{2}\;\sigma^{2}_{\xi}\;+\;
\left(\frac{\partial \log\epsilon}{\partial  \mathrm{[Fe/H]}}\right)^{2}\sigma^{2}_{\mathrm{[Fe/H]}} \;+\; \\ 2\bigg [\left(\frac{\partial \log\epsilon}{\partial T}\right) \left(\frac{\partial \log \epsilon}{\partial \log g}\right) \sigma_{T,\log g}
\;+\; \left(\frac{\partial \log\epsilon}{\partial \xi}\right) \left(\frac{\partial \log \epsilon}{\partial \log g}\right) \sigma_{\log g, \xi} \\
\;+\;\left(\frac{\partial \log\epsilon}{\partial \xi}\right) \left(\frac{\partial \log \epsilon}{\partial T}\right) \sigma_{ \xi, T}\Bigg],
\end{split}
\end{equation}
\noindent where $\sigma_{T}$, $\sigma_{\log g}$, and $\sigma_{\xi}$ are the typical uncertainties on the atmospheric parameters. These values are estimated as $\sigma_{T}$ = 34~K, $\sigma_{\log g}$ = 0.6~dex, and $\sigma_{\xi}$ = 0.05~km/s. The uncertainty on metallicity was estimated as  $\sigma_{\mathrm{[Fe/H]}}$ = 0.12~dex.
The partial derivatives that appear in Eq.~\ref{Eq:Johnson} were evaluated by varying the atmospheric parameters $T_{\rm eff}$, $\log g$, microturbulence ($\xi$), and [Fe/H] by 100~K, 0.5, 0.5~km/s, and 0.5 dex, respectively.
The covariances $\sigma_{T,\log g}$, $\sigma_{\log g,\xi}$, and $\sigma_{\xi, T}$ were derived using the same method as given by \cite{Johnson2002}. In order to calculate  $\sigma_{T,\log g}$, we varied the temperature while fixing metallicity and microturbulence, and determined the $\log g$ value required for ensuring the ionisation balance. Then, using Eq.~3 of \cite{Johnson2002}, we derived the covariance, $\sigma_{T,\log g}$, and find a value of 23. Similarly, we find $\sigma_{\log g,\xi}$ = 0.01 and $\sigma_{\xi, T}$ = 83. 

The random error, $\sigma_{\rm ran}$, is the line-to-line scatter. For elements for which we used more than four lines to derive the abundances, we adopted $\sigma_\mathrm{ran} = \sigma_{l}/N^{1/2}$, where $\sigma_{l}$ is the standard deviation of the abundances derived from all the $N$ lines of the considered element. For the elements for which fewer numbers of lines were used to derive the abundances, we selected a $\sigma_{\rm ran}$ value as described in \cite{Karinkuzhi-2021}. The final error on [X/Fe] was derived as

\begin{equation}
\sigma^{2}_{\rm [X/Fe]} =\sigma^{2}_{\rm X} + \sigma^{2}_{\rm Fe} - 2\;\sigma_{\rm X,Fe},
\end{equation}\

\noindent where $\sigma_{\rm X,Fe}$ was calculated using Eq.~6 from \cite{Johnson2002} with the additional term $\left(\frac{\partial \log\epsilon}{\partial  \mathrm{[Fe/H]}}\right)$ included.

\section{The abundance distance, $d$}\label{sec:d}

In order to characterise in more detail the abundance profile of 56 UMa, we computed an abundance distance from the solar r-process abundance pattern following \cite{Karinkuzhi-2021}. We used seven chemical elements (Y, Zr, Ba, La, Ce, Nd, and Sm), instead of only two as in \cite{Karinkuzhi-2021}, to compute the abundance distance from the solar r-process abundance pattern. We also used Eu as the normalising element between the reference r-process abundance profile and the abundance profile of our target, because it is formed mainly via the r-process and is easily measurable in most stars.

The distance, $d$, was defined as follows by \cite{Karinkuzhi-2021}:

\begin{equation}
d = \frac{1}{N}\sum_{x_i} (\log_{10} \epsilon_{x_i,\ast} - \log_{10} \epsilon_{x_i,\rm{norm(r,\ast)}}),
\label{Eq:dist-signed}
\end{equation}

\noindent where $\{x_1... x_N\}$ is the list of the $N$ considered heavy elements, and we use the usual notation $\log_{10} \epsilon_{x_i} = \log_{10} (n_{x_i}/n_{\rm H}) + 12$, with $n_{x_i}$ the number density of element $x_i$. We denote as $\log_{10} \epsilon_{x_i,\ast}$ the abundance of element $x_i$ as measured for 56~UMa (Table~\ref{Tab:abundances}), and $\log_{10} \epsilon_{x_i,\rm{norm(r,\ast)}}$, the standard r-process abundance profile $\log \epsilon_{x_i, r}$  normalised to the star abundance profile with respect to europium:

\begin{equation}
\log \epsilon_{x_i,\rm{norm(r,\ast)}} = \log \epsilon_{x_i, r} + (\log\epsilon_{Eu,\ast} - \log\epsilon_{Eu,r}).
\end{equation}

\noindent The adopted r-process abundances, $\log \epsilon_{x_i,r}$, are listed in \cite{Karinkuzhi-2021}. We find $d = 0.54$ for 56~UMa.

When adopting instead the RMS distance as defined in \cite{Karinkuzhi-2021},
\begin{equation}
d_{\rm RMS} = \left( \frac{1}{N}\sum_{x_i} (\log_{10} \epsilon_{x_i,\ast} - \log_{10} \epsilon_{x_i,\rm{norm(r,\ast)}})^2 \right)^{1/2},
\label{Eq:dist-RMS}
\end{equation}
we find $d_{\rm RMS} = 0.63$, well within the bulk of the RMS distances of hybrid-profile stars such as CEMP-rs objects.

\section{Exploding stars in binary systems}\label{sec:kick}

We followed \cite{Postnov2014} and used the fact that the SN explosion in 56~UMa did not unbind the system to set constraints on the initial masses. We adopt the subscript `p' for the current primary star, the giant, and `s' for the NS and its progenitor. Assuming an initially circular orbit, a product of strong tidal circularisation, and a symmetric explosion (no kick imparted on the NS), the mass ejected by the explosion, $\Delta M_{\rm s}$, must fulfil the following relation for the system to remain bound:
\begin{equation}
\label{Eq:bound}
 \Delta M_{\rm s} < 0.5\;(M_{\rm s,i}+M_{\rm p,i}). 
\end{equation}
This expression, only valid for a symmetric explosion, is the equivalent to imposing that the post-explosion eccentricity,
\begin{equation}
\label{Eq:e_sym}
    e = \frac{(M_{\rm p,i} + M_{\rm s,i}) - (M_{\rm p,f} + M_{\rm s,f})}{M_{\rm p,f} + M_{\rm s,f}}
\end{equation}\
is smaller than unity (Eq.~51 in \cite{Postnov2014}).

\cite{Schmelz2022} suggest that the red giant has ejected 1.2~\Msun\ since the explosion of its companion via wind mass loss (visible as a hydrogen feature close to 56~UMa in Fig. \ref{cavity}). Hence, plugging $M_{\rm p,f} = 5.5$~\Msun\ (the currently measured 4.3~\Msun\ plus the 1.2~\Msun\ ejected via wind mass loss), $M_{\rm s,f} = 1.3$~\Msun, and the measured eccentricity of 0.562 (Table \ref{table:MCMCresults}) into Eq.~\ref{Eq:e_sym}, we obtain $\Delta M_{\rm s} = 3.8$~\Msun, which is not enough considering that a star needs to have a mass of at least 8~\Msun\ to explode as a SN.

Since the discrepancy clearly indicates that the above picture is far too simple, we considered how an asymmetric SN explosion and its associated kick can modify the orbital eccentricity of a system. Again, according to \cite{Postnov2014} (Eqs. 46-47),
\begin{equation}
\label{Eq:e_kick}
    1-e^2 = \chi\frac{a_{\rm i}}{a_{\rm f}}\left(\frac{w_z^2+(V_i+w_y)^2)}{V_i^2}\right) 
\end{equation}
with 
\begin{equation}
\label{Eq:a}
    \frac{a_{\rm i}}{a_{\rm f}} = \left[2-\chi\left(\frac{w_x^2+w_z^2+(V_i+w_y)^2)}{V_i^2}\right)\right],
\end{equation}\\

\noindent where $\chi = (M_{\rm p,i} + M_{\rm s,i})/(M_{\rm p,f} + M_{\rm s,f})$. The kick velocity imparted to the NS is expressed as $(w_x, w_y, w_z)$ in an instantaneous reference frame centred on the primary star (the companion to the exploding star), with the $x$-axis directed from the primary to the secondary, the $y$-axis pointing in the direction of the relative pre-SN orbital velocity, $\mathbf{V}_i$, of the primary around the secondary (the NS star), and the $z$-axis perpendicular to the orbital plane. In this frame, the pre-SN relative velocity is $\mathbf{V}_i = (0, V_{\rm i}, 0)$, where $V_{\rm i} = \left(G (M_{\rm p,i} + M_{\rm s,i})/a_i\right)^{1/2}$. 

It is easy to show that if the kick imparted to the NS by the asymmetric explosion is along the orbital motion, thus of the form $(0, w_y, 0)$, the post-explosion eccentricity is
\begin{equation}
\label{Eq:e_kick_2}
e = \frac{\xi^2 M_{\rm tot,i} - M_{\rm tot,f}}{M_{\rm tot,f}}, 
\end{equation}
where $M_{\rm tot,i}$ and $M_{\rm tot,f}$ are the initial and final total mass in the system, and with $\xi = (V_{\rm i}+w_y)/V_{\rm i}$. Thus, a kick with $w_y < 0$ and $w_x = w_z = 0$ has the same effect as reducing the effective initial mass by a factor $\xi^2$.

As discussed in Sect. \ref{sec:discussion}, the evolutionary timescales give important constraints. The two stars must have had similar initial masses for the current giant to be a giant. If it ejected 1.2~\Msun\ since the explosion, and the stars were initially of almost equal mass, there would be an extra 2~\Msun\ that had to have been blown away by the passing SN remnant \citep{Hirai2014}. 

In summary, we reach the following picture: $M_{\rm p,f} = 5.5$~\Msun\ for the giant companion just after the explosion (currently 4.3~\Msun) and $M_{\rm s,f} =  1.3$~\Msun\ for the NS, or $M_{\rm tot,f} = 6.8$~\Msun. Along with the eccentricity of 0.562, this yields $\xi^2 M_{\rm tot,i} = 1.562 \; M_{\rm tot,f} = 10.62$~\Msun. If $M_{\rm s,i} =  8$~\Msun\ and $M_{\rm p,i} =  7.5$~\Msun,  $M_{\rm tot,i} =  15.5$~\Msun, so $\xi^2 = 0.685$ or $w_y = -0.17 \; V_{\rm i} \sim 6$~km/s, a value compatible with simulations of electron-capture SNe \citep[e.g.][]{Gessner18}. With these values, Eq.~\ref{Eq:a} moreover predicts $a_{\rm i} = 10$~AU, thus leaving enough room for the NS progenitor to reach the SN stage without filling its Roche lobe.

Finally, we wanted to estimate if the energy of the explosion was enough to unbind 2~\Msun. If at the time of the explosion the core of the giant star had a radius of 58~R$_\odot$, then the energy needed to unbind 2~\Msun\ from such a giant star would be $\sim 7.2\; 10^{47}$~erg, of the order of the energy budget of a typical type II SN ($10^{51}$~erg, translated to $7.2\;10^{47}$~erg considering the dilution factor $(R_g/a_i)^2$ with $R_g \sim 58$~R$_\odot$ and $a_i \sim 10$~au, according to Eq.~\ref{Eq:a}).

\end{document}